\documentclass[journal]{IEEEtran}
\ifCLASSINFOpdf
\else
\fi
\usepackage{cite}
\usepackage{amssymb}
\usepackage{mathtools}
\usepackage{fancyhdr}
\usepackage{algorithm}
\usepackage{algpseudocode}
\usepackage{graphicx}
\usepackage{color}

\begin{document}
%
\title{Fronthaul Compression and Beamforming Optimization for Secure Cell-free ISAC Systems}
%
%
%

\author{Seongjun Kim and Seongah Jeong,~\IEEEmembership{Senior Member,~IEEE}
\thanks{This work was supported by the National Research Foundation of Korea(NRF) grant funded by the Korea government (MSIT) (No. 2023R1A2C2005507)
}
\thanks{Seongjun Kim is with the School of Electronics Engineering, Kyungpook National University, Daegu 14566, South Korea (e-mail: ksj6989@knu.ac.kr).}
\thanks{Seongah Jeong is with the School of Advanced Fusion Studies, Department of Intelligent Semiconductor Engineering, University of Seoul, Seoul 02504, South Korea (e-mail: seongah@uos.ac.kr).}}

%
%

\markboth{}%
{Shell \MakeLowercase{\textit{et al.}}: Bare Demo of IEEEtran.cls for IEEE Journals}
%



\maketitle

\begin{abstract}
This letter aims to provide sensing capabilities for a potential eavesdropper, while simultaneously enabling the secure communications with the legitimate users in a cell-free multiple-input multiple-output system with limited fronthaul links. In order to maximize the sensing performance, the joint design of fronthaul compression and beamforming is proposed considering the constraints on the finite fronthaul-capacity links and the maximum power along with the worst-case secrecy rate requirements. To this end, we propose an algorithmic solution based on the minorization-maximization method and semidefinite programming relaxation techniques, whose performance superiority is verified via simulations compared to the reference schemes such as distributed sensing and random beamforming.
\end{abstract}


%
\IEEEpeerreviewmaketitle

\section{Introduction}
%
%
%
%
\IEEEPARstart{I}{ntegrated} sensing and communication (ISAC) technology is expected to be widely used in various applications such as eXtended reality (XR), autonomous vehicles, navigation, and other location-based services. However, by sharing the spectrum resources for both communications and sensing, the transmit signal needs to be steered toward the target as well as the legitimate communication users to enhance sensing performance, which might lead to the information leakage to the potential eavesdropper (Eve). To mitigate the privacy issue, the physical-layer security technologies using the various multiple-antenna techniques have been actively explored for ISAC, which are referred to as the secure ISAC systems \cite{r2, r3, r4, r5}. In \cite{r2}, a single ISAC transmitter is considered for communicating with a single user and detecting multiple targets, addressing the problem of minimizing the beampattern matching errors under a secrecy rate constraint. The authors in \cite{r3} propose the MIMO radar beampattern to minimize the signal-to-noise ratio (SNR) at a potential Eve with the perfect and imperfect channel state information (CSI).  Most of the existing studies \cite{r2, r3} on the secure ISAC have focused on a single-cell scenario, which need to be extended to multi-cell scenarios to manage the inter-cell interference effectively.

\begin{figure}[t]
    \centering
    \includegraphics[width= 7.5cm]{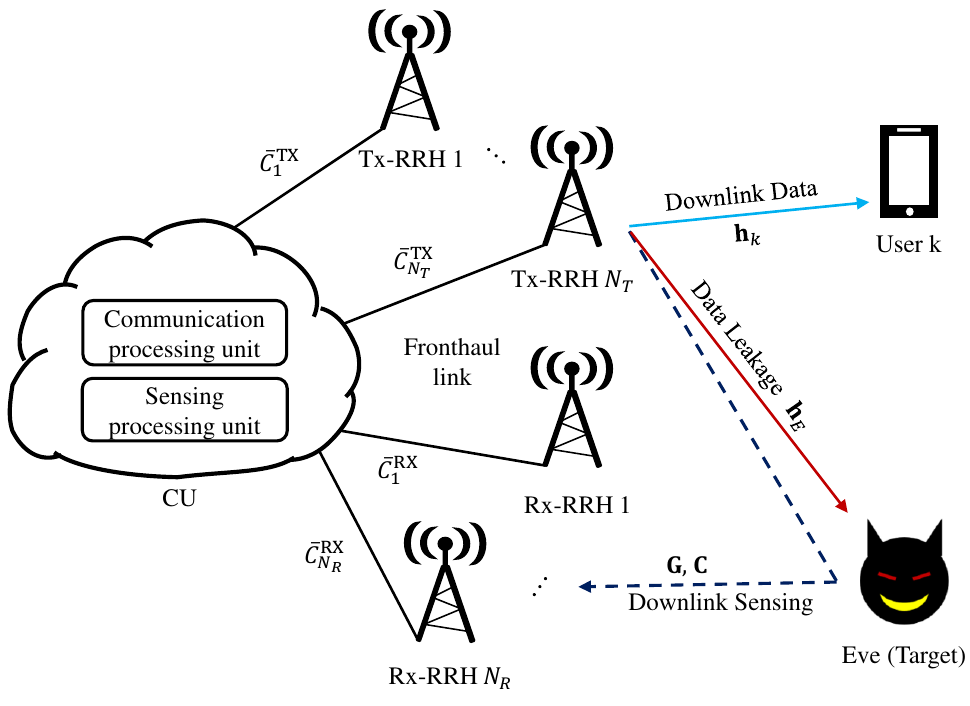} 
    \caption{A secure cell-free ISAC system to simultaneously serve the downlink data communication and downlink passive sensing.}
    \vspace{-0.4cm}
    \label{fig:fig1}
\end{figure}

Recently, the secure cell-free ISAC has been investigated to enhance both communication and sensing performances through the joint signal processing of the multiple remote radio heads (RRHs) \cite{r4, r5, r16}. In \cite{r5}, the authors address an optimization problem to maximize the target detection probability, while considering signal-to-interference-plus-noise-ratio (SINR) for communication users, SNR for information eavesdroppers, and the detection probability for sensing eavesdroppers. In \cite{r16}, the  secure spatial signal design to minimize the Cram\'{e}r-Rao Bound as the sensing performance metric is studied to meet the SINR and SNR requirements of the desired users and Eve, respectively. The recent studies \cite{r4, r5, r16} on secure cell-free ISAC systems mostly assume the ideal fronthaul links between a central unit (CU) and RRHs that is not possible in real applications. For the effective collaboration among the RRHs, the  functional split between CU and RRHs needs to be discussed under constrained fronthaul capacity.

By these motivations, we consider a secure cell-free ISAC system for serving the downlink data communications to the multiple users with the finite-capacity fronthaul links, while providing sensing capability for a single Eve.
With the aim of maximizing the sensing performance, we propose the joint design of the transmit beamforming and the fronthaul quantization subject to constraints on the secrecy rate, transmit power and fronthaul capacity. To this end, we develop an algorithmic solution based on minorization-maximization (MM) method and a semidefinite programming (SDP) relaxation technique, whose performance superiority is validated via simulations in comparison with the benchmark schemes of distributed sensing or random beamforming.

\section{SYSTEM MODEL}
We consider a secure cell-free ISAC system as shown in Fig. 1, consisting of a CU, $N_T$ transmit RRHs (Tx-RRHs), $N_R$ receive RRHs (Rx-RRHs), $K$ users, and a passive Eve. The CU provides the downlink secure communications and at the same time attempts to sense the Eve as a sensing target. The sets of Tx-RRHs, Rx-RRHs and users are denoted as $\mathcal{N}_t$$=$$\{1, ..., N_T\}$, $\mathcal{N}_r$$=$$\{1,...,N_R\}$ and $\mathcal{K}$$=$$\{1,...,K\}$, respectively. The Tx-RRHs transmit the cooperative data signals and the Rx-RRHs receive the sensing echoes reflected by the Eve. Here, for the processing, the signals are compressed and forwarded to the CU via constrained fronthaul links. We assume that all RRHs are fully synchronized \cite{r3, r4}, each of which is equipped with a uniform linear array (ULA) with $N_A$ antenna elements and connected to the the CU via the orthogonal finite-capacity fronthaul links. The $K$ ground users and Eve are assumed to have a single antenna. The CU and the users have the full CSI for their respective channels including the Eve that can be obtained from the past experiences, e.g., via camera or sensing \cite{r4}. 

We denote the data symbol of user $k$$\in$$\mathcal{K}$ as $s_k[m]$ at the $m$th symbol time and define the collection of all symbol data of users as $\mathbf{s}[m]$$\triangleq$$[s_{1}[m] , \ldots , s_{K}[m]]^T$, whose entries are independent and identically distributed (i.i.d) complex Gaussian random variables, denoted as $\mathcal{CN}(0, 1)$. In the centralized design, we represent the precoded data signal $\mathbf{W}_i\mathbf{s}[m]$ of Tx-RRU $i$, where $\mathbf{W}_i$$=$$[\mathbf{w}_{i, 1},\ldots,\mathbf{w}_{i, K}]$$\in$$\mathbb{C}^{N_A \times K}$ represents the beamforming matrix of Tx-RRU $i$ that collects all the beamforming vector $\mathbf{w}_{i, k}$$\in$$\mathbb{C}^{N_A \times 1}$ for all $k$$\in$$\mathcal{K}$. Due to the finite-capacity fronthaul link between Tx-RRU $i$ and CU, the CU quantizes the baseband signal $\tilde{\mathbf{x}}_i$ and forwards it to the Tx-RRU $i$. Accordingly, the received quantized signal at Tx-RRU $i$ is written as 
$\mathbf{x}_{i}[m]$$=$$\tilde{\mathbf{x}}_{i}[m]$$+$$ \mathbf{q}_{i}^{\mathrm{TX}
}$$=$$\mathbf{W}_{i}\mathbf{s}[m]$$+$$\mathbf{q}_{i}^{\mathrm{TX}},
\label{eq:communication signal}$
where the quantization noise vector $\mathbf{q}_{i}^{\mathrm{TX}}$ is assumed to have i.i.d. complex Gaussian entries with zero-mean and covariance matrix $\mathbf{Q}_i^{\mathrm{TX}}$$=$$\sigma^2_{\mathrm{TX}, i}\mathbf{I}$ and is independent for different $i$$\in$$\mathcal{N}_t$. With the independence of $\mathbf{q}_i^{\mathrm{TX}}$, the transmit power of the $i$th RRH can be expressed as $P_{i}(\mathbf{W}_{i}, \mathbf{Q}_{i}^{\mathrm{TX}})$$=$$ \mathbb{E}\{\|\mathbf{x}_{i}[m]\|^2\}$$=$$\mathrm{tr}(\mathbf{W}_{i}\mathbf{W}_{i}^H + \mathbf{Q}_{i}^{\mathrm{TX}})$,
which needs to satisfy the maximum transmit power constraint $\bar{P}_i$, i.e., $P_{i}(\mathbf{W}_{i}, \mathbf{Q}_{i}^{\mathrm{TX}}) \leq \bar{P}_i$. 

In the downlink communications, the received signals at user $k \in \mathcal{K}$ is expressed as
\vspace{-0.2cm}
\begin{eqnarray}
&&\hspace{-1.2cm}y_k[m] = \sum_{i=1}^{N_T}\mathbf{h}_{k,i}^H\mathbf{x}_i[m] + n_k[m] = \mathbf{h}_k^H \mathbf{w}_{k}s_{k}[m] \nonumber\\
&&+\hspace{-0.2cm}\sum_{k'\in \mathcal{K} \backslash k} \hspace{-0.2cm}\mathbf{h}_k^H \mathbf{w}_{k'}s_{k'}[m]+\mathbf{h}_k^H\mathbf{q}[m] + n_k[m], 
\end{eqnarray}
where the vector $\mathbf{h}_{k,i}^H$ denotes the channel between Tx-RRH $i$ and user $k$, whose collection for user $k$ is defined as $\mathbf{h}_k$$\triangleq$$[\mathbf{h}_{k,1}^T, \ldots, \mathbf{h}_{k,N_{\mathrm{t}}}^T]^T$$\in$$\mathbb{C}^{N_T N_A \times 1}$. The downlink channel is modeled using Rician fading, expressed as $\mathbf{h}_{k,i}$$=$$\sqrt{R/(R+1)}\mathbf{f}_{k,i}$$+$$ \sqrt{1/(R+1)}\mathbf{\tilde{f}}_{k,i}$, where the line-of-sight (LOS) component is represented as $\mathbf{f}_{k,i}$$=$$[1, \exp(-j\pi\sin(\theta)), \ldots, \exp(-j\pi(N_A - 1)\sin(\theta))]^T \in \mathbb{C}^{N_A \times 1}$, while the non-line-of-sight (NLOS) component, $\mathbf{\tilde{f}}_{k,i}$, is modeled as a circularly symmetric complex Gaussian (CSCG) random variable, with a variance of 1. The vector $\mathbf{w}_k$$\triangleq$$[\mathbf{w}_{k,1}^T, \ldots, \mathbf{w}_{k,N_{\mathrm{t}}}^T]^T$$\in$$\mathbb{C}^{N_T N_A \times 1}$ is the beamforming vector for user $k$ at all Tx-RRHs, the vector $\mathbf{q}[m]$ is collection of the quantization noise vectors $\mathbf{q}[m]$$=$$ [\mathbf{q}_1^T, \ldots, \mathbf{q}_{N_T}^T]^T$ for all $i$$\in$$\mathcal{N}_t$, and $n_k[m]$ is the additive white Gaussian noise (AWGN) with zero-mean and variance $\sigma_{k}^2$ for all $k$$\in$$ \mathcal{K}$.

For sensing the Eve, we consider the downlink passive sensing, where Rx-RRHs receive echoes reflected by Eve and forward them to the CU for detection. 
The sensing channels are assumed to consist of target scattering effects and clutter components. The target scattering effects between Tx-RRH $i$ and Rx-RRH $j$, scattered by the Eve, can be modeled as $\mathbf{G}_{j,i}$$=$$\alpha_{j,i} \pmb{a}(\theta_j) \pmb{a}^T(\theta_i)$ \cite{r4}, where $\alpha_{j,i}$$\sim$$\mathcal{CN}(0, \sigma_{G, j,i}^2)$ represents the combined effects of the bi-static radar cross-section (RCS) of the Eve and the path loss effect, the vectors $\pmb{a}(\theta)$$=$$[1, \exp(-j\pi\sin(\theta)), \ldots, \exp(-j\pi(N_A - 1)\sin(\theta))]^T$$\in$$\mathbb{C}^{N_A \times 1}$ denotes the array response vectors, and $\theta_i$ and $\theta_j$ are angle of departure (AoD) of Tx-RRH $i$ and angle
of arrival (AoA) of Rx-RRH $j$. The received signal at Rx-RRH $j$ is then expressed as
\vspace{-0.1cm}
\begin{eqnarray} 
    \mathbf{r}_{j}[m] = \sum_{i=1}^{N_T}\mathbf{G}_{j,i}\mathbf{x}_i[m] + \sum_{i=1}^{N_T} \mathbf{C}_{j,i} \mathbf{x}_i[m] + \mathbf{n}_j[m],
\label{eq:sensing signal}
\end{eqnarray}
where $\mathbf{C}_{j,i}$ represents the clutter matrix with the i.i.d $\mathcal{CN}(0, \sigma_{C, j}^2)$ entries \cite{r10}, and $\mathbf{n}_j[m]$$\sim$$\mathcal{CN}(\mathbf{0}, \sigma_{N, j}^2\mathbf{I}_{N_A})$ is the AWGN of Rx-RRH $j$. For the centralized sensing at the CU, the signal $\mathbf{r}_j[m]$ at Rx-RRH $j$ in (\ref{eq:sensing signal}) is compressed and forwarded to the CU via the fronthaul link. Accordingly, the received sensing signal at CU is given as $\tilde{\mathbf{r}}_{j}[m]$$=$$\mathbf{r}_{j}[m]$$+$$ \mathbf{q}^{RX}_j[m]$, where the quantization noise vector $\mathbf{q}^{RX}_j[m]$ follows a zero-mean Gaussian distribution with the covariance matrix $\mathbf{Q}^{RX}_j$$=$$\sigma^2_{\mathrm{RX}, j}\mathbf{I}$, i.e., $\mathbf{q}^{RX}_j[m]$$\sim$$\mathcal{CN}(\mathbf{0},\mathbf{Q}_j^{\mathrm{RX}})$, and is independent for different $j$$\in$$\mathcal{N}_r$. Based on the overall received sensing signals from Rx-RRHs, the CU attempts to detect the Eve by using the binary hypothesis test with an optimum detector \cite{r10}. In this process, the sensing performance depends on the SINR of the collected sensing signals at the CU, which is associated with the sensing ability such as the detection/false alarm probability or Cram\'{e}r-Rao Bound \cite{r4, r5}.
For this reason, we adopt the sensing SINR as the sensing performance metric, defined as the ratio between the power from the signal of target scattering effects and the combined power from the signals of clutter, noise and quantization noise. The sensing SINR at CU is calculated as
\begin{eqnarray} 
&&\hspace{-0.7cm}\gamma_s(\hspace{-0.05cm}\mathbf{W}\hspace{-0.05cm}, \mathbf{Q}^{\mathrm{TX}}\hspace{-0.05cm}, \hspace{-0.05cm}\mathbf{Q}^{\mathrm{RX}}\hspace{-0.05cm}) \nonumber\\
&&\hspace{-0.7cm}=\hspace{-0.05cm} \frac{\mathrm{tr}(\hspace{-0.05cm}\mathbf{G}(\hspace{-0.05cm}\mathbf{W}\mathbf{W}^H\hspace{-0.1cm}+\hspace{-0.05cm}\mathbf{Q}^{\mathrm{TX}}\hspace{-0.05cm})\mathbf{G}^H\hspace{-0.05cm})}{\mathrm{tr}(\hspace{-0.05cm}\mathbf{C}(\hspace{-0.05cm}\mathbf{W}\mathbf{W}^H\hspace{-0.1cm}+\hspace{-0.05cm}\mathbf{Q}^{\mathrm{TX}}\hspace{-0.05cm})\mathbf{C}^H\hspace{-0.05cm})\hspace{-0.05cm} +\hspace{-0.05cm}\sum_{j=1}^{N_R}\hspace{-0.05cm}N_A \sigma_{y,j}^2\hspace{-0.05cm}+\hspace{-0.05cm}\sum_{j=1}^{N_R}\hspace{-0.05cm}\mathrm{tr}(\hspace{-0.05cm}\mathbf{Q}^{\mathrm{RX}}_j\hspace{-0.05cm})},\nonumber\\
\label{eq:sensing SINR}
\end{eqnarray}
where the matrix $\mathbf{G}$$=$$[\mathbf{G}_1^T, \ldots, \mathbf{G}_{N_R}^T]^T$ is a block matrix composed of $\mathbf{G}_j$$=$$[\mathbf{G}_{j, 1}, \ldots, \mathbf{G}_{j, N_T}]$, and each $\mathbf{G}_{j,i}$ represents the scattering effects between Tx-RRH $i$ and Rx-RRH $j$, the matrix $\mathbf{W}$$=$$[\mathbf{W}_1^T,\ldots,\mathbf{W}_{N_T}^T]^T$ is the collection of transmit beamforming matrices $\{\mathbf{W}_i\}_{i\in \mathcal{N}_t}$, and the matrix $\mathbf{C}$$=$$[\mathbf{C}_1^T, \ldots, \mathbf{C}_{N_R}^T]^T$ represents the matrix to collect all the clutter components with $\mathbf{C}_j$$=$$[\mathbf{C}_{j, 1}, \ldots, \mathbf{C}_{j, N_T}]$.

For secure communication of users, we define the SINR of the $k$th user as 
\vspace{-0.2cm}
\begin{eqnarray} 
&&\hspace{-1.5cm}\gamma_k(\mathbf{W}, \mathbf{Q}^{\mathrm{TX}}) \nonumber\\
&&\hspace{-1cm}= \frac{|\mathbf{h}_k^H \mathbf{w}_k|^2}{\sum_{k'\in \mathcal{K} \backslash k} |\mathbf{h}_k^H \mathbf{w}_{k'}|^2 + \mathbf{h}_k^H\mathbf{Q}^{\mathrm{TX}}\mathbf{h}_k + \sigma_k^2},  \forall k \in \mathcal{K},
\end{eqnarray}
where $\mathbf{Q}^{\mathrm{TX}}$$=$$\mathrm{diag}(\mathbf{Q}_1^{\mathrm{TX}}, \ldots, \mathbf{Q}_{N_T}^{\mathrm{TX}})$ due to the independence of ${\mathbf{q}_i^{\mathrm{TX}}}$. The leakage signal at Eve for user $k$$\in$$\mathcal{K}$ is denoted as $y_E[m]$$=$$\sum_{k'\in \mathcal{K} \backslash k}\mathbf{h}_E^H \mathbf{w}_{k'}s_{k'}[m]+\mathbf{h}_E^H\mathbf{q}[m] + n_E[m]$, where $\mathbf{h}_E$$\triangleq$$[\mathbf{h}_{E,1}^T, \ldots, \mathbf{h}_{E,N_{\mathrm{t}}}^T]^T$$\in$$\mathbb{C}^{N_T N_A \times 1}$ denotes leakage channel between the Eve and Tx-RRHs. Similar to the users' communication channels, the leakage channel is also modeled as a Rician channel. The SINR of the leakage signal is given as $\gamma_E(\mathbf{W}, \mathbf{Q}^{\mathrm{TX}})$$=$$|\mathbf{h}_E^H \mathbf{w}_k|^2/(\sum_{k'\in \mathcal{K} \backslash k}\hspace{-0.05cm}|\mathbf{h}_E^H \mathbf{w}_{k'}|^2\hspace{-0.05cm} + \hspace{-0.05cm}\mathbf{h}_E^H\mathbf{Q}^{\mathrm{TX}}\mathbf{h}_E \hspace{-0.04cm}+\hspace{-0.04cm}\sigma_E^2\hspace{-0.04cm})$, where $\sigma_E^2$ denotes the variance of AWGN at Eve. Therefore, the worst-case secrecy rate is expressed as \cite{r2}
\vspace{-0.2cm}
\begin{eqnarray} 
&&\hspace{-1cm}R_{\mathrm{Sec}}(\mathbf{W}, \mathbf{Q}^{\mathrm{TX}}) \nonumber\\
&&\hspace{-1cm}= \min_k\hspace{-0.05cm}\left[ \log\hspace{-0.05cm}\left(1 + \gamma_k(\hspace{-0.05cm}\mathbf{W}\hspace{-0.05cm}, \mathbf{Q}^{\mathrm{TX}}\hspace{-0.05cm})\hspace{-0.05cm}\right) 
\hspace{-0.05cm} -\hspace{-0.05cm} \log\hspace{-0.05cm}\left(1 \hspace{-0.05cm}+ \hspace{-0.05cm}\gamma_E(\hspace{-0.05cm}\mathbf{W}\hspace{-0.05cm}, \mathbf{Q}^{\mathrm{TX}})\hspace{-0.05cm}\right) \hspace{-0.05cm}\right]^+\hspace{-0.05cm},
\end{eqnarray}
where $[a]^+$$=$$\max(a, 0)$.

Moreover, since the limited fronthaul links between CU and RRHs are considered, the fronthaul design needs to satisfy the fronthaul capacity constraint. Using the rate-distortion theory \cite{r7}, the required fronthaul rates between the CU and Tx-RRH/Rx-RRH are given as
\vspace{-0.2cm}
\begin{eqnarray}
&&\hspace{-0.7cm}R_{i}^{\mathrm{TX}}(\mathbf{W}_{i}, \mathbf{Q}_{i}^{\mathrm{TX}}) \nonumber\\
&&=\log\hspace{-0.05cm}\det(\mathbf{W}_i\mathbf{W}_i^H + \mathbf{Q}_{i}^{\mathrm{TX}})-\log\hspace{-0.05cm}\det(\mathbf{Q}_{i}^{\mathrm{TX}}) \quad\text{and} \\
&&\hspace{-0.7cm}R_{j}^{\mathrm{RX}}\hspace{-0.05cm}(\hspace{-0.05cm}\mathbf{W}\hspace{-0.05cm},\hspace{-0.05cm} \mathbf{Q}^{\mathrm{TX}}\hspace{-0.05cm}, \hspace{-0.05cm}\mathbf{Q}^{\mathrm{RX}}\hspace{-0.05cm}) \hspace{-0.05cm}= \hspace{-0.05cm}\log\hspace{-0.05cm}\det(\mathbb{E}[\mathbf{r}_{j}[m]\mathbf{r}_{j}[m]^H])\hspace{-0.05cm}-\hspace{-0.05cm}\log\hspace{-0.05cm}\det(\hspace{-0.05cm}\mathbf{Q}_j^{\mathrm{RX}}\hspace{-0.05cm}) \nonumber\\
&&= \log\hspace{-0.05cm}\det((\hspace{-0.05cm}\mathbf{G}_j\hspace{-0.05cm}+\hspace{-0.05cm}\mathbf{C}_j\hspace{-0.05cm})\hspace{-0.05cm}(\hspace{-0.05cm}\mathbf{W}\mathbf{W}^H \hspace{-0.1cm}+\hspace{-0.05cm} \mathbf{Q}^{\mathrm{TX}}\hspace{-0.05cm})(\hspace{-0.05cm}\mathbf{G}_j\hspace{-0.05cm}+\hspace{-0.05cm}\mathbf{C}_j\hspace{-0.05cm})^H \nonumber\\
&&\quad+\sigma_{y,j}^2\mathbf{I}_{N_A}\hspace{-0.05cm}+\hspace{-0.05cm}\mathbf{Q}_j^{\mathrm{RX}})\hspace{-0.05cm}-\hspace{-0.05cm}\log\hspace{-0.05cm}\det(\mathbf{Q}_j^{\mathrm{RX}}),
\end{eqnarray}
respectively.
\section{PROBLEM FORMULATION AND PROPOSED ALGORITHM}
In this paper, we aim to maximize the sensing SINR subject to the constraints on two types of fronthaul links, transmit power and the worst-case secrecy rate of users by jointly optimizinig the transmit beamforming matrix $\mathbf{W}$ and the quantization noise covariances $\mathbf{Q}^{\mathrm{TX}}$ and $\mathbf{Q}^{\mathrm{RX}}$. To this end, the optimization problem is formulated as
\vspace{-0.2cm}
\begin{subequations}\label{p1}
\begin{eqnarray}
\hspace{-0.5cm}\underset{\mathbf{W}, \mathbf{Q}^{\mathrm{TX}}, \mathbf{Q}^{\mathrm{RX}}}{\mathrm{max}}&&\hspace{-0.5cm}\gamma_s(\mathbf{W},\hspace{-0.05cm}\mathbf{Q}^{\mathrm{TX}},\hspace{-0.05cm} \mathbf{Q}^{\mathrm{RX}}),\\
\text{s.t.}\qquad&&\hspace{-0.5cm}R_{\mathrm{Sec}}(\mathbf{W},\hspace{-0.05cm} \mathbf{Q}^{\mathrm{TX}})\hspace{-0.05cm}\geq\hspace{-0.05cm} \bar{R}_{\mathrm{Sec}},\\
&&\hspace{-0.5cm}R_{i}^{\mathrm{TX}}(\mathbf{W}_{i},\hspace{-0.05cm} \mathbf{Q}^{\mathrm{TX}}_{i})\hspace{-0.05cm}\leq\hspace{-0.05cm} \bar{C}_i^{\mathrm{TX}},\hspace{+0.05cm} \forall i \in \mathcal{N}_t,\\
&&\hspace{-0.5cm}R_{j}^{\mathrm{RX}}(\mathbf{W},\hspace{-0.05cm} \mathbf{Q}^{\mathrm{TX}},\hspace{-0.05cm} \mathbf{Q}_j^{\mathrm{RX}})\hspace{-0.05cm}\leq\hspace{-0.05cm} \bar{C}_j^{\mathrm{RX}},\hspace{+0.05cm} \forall j \in \mathcal{N}_r,\\
&&\hspace{-0.5cm}P_{i}(\mathbf{W}_{i},\hspace{-0.05cm} \mathbf{Q}^{\mathrm{TX}}_{i})\hspace{-0.05cm}\leq\hspace{-0.05cm} \bar{P}_i,\hspace{+0.05cm} \forall i \in \mathcal{N}_t,
\end{eqnarray}
\end{subequations}
where $\bar{R}_{\mathrm{Sec}}$ indicates the required secrecy rate for the users, $\bar{C}_i^{\mathrm{TX}}$ and $\bar{C}_j^{\mathrm{RX}}$ denote the fronthaul link capacities from Tx-RRHs $i$ and Rx-RRHs $j$ to the CU, respectively, and $\bar{P}_i$ represents the maximum transmit power available at Tx-RRH $i$. To tackle this problem, we first define the covariance transmit beamforming matrices $\mathbf{V}_{k}$$\triangleq$$\mathbf{w}_{k}\mathbf{w}_k^H$ and $\mathbf{V}_i$$=$$\mathbf{W}_{i}\mathbf{W}_i^H$. Then, the problem (\ref{p1}) can be reformulated as
\vspace{-0.2cm}
\begin{subequations}\label{p2}
\begin{eqnarray}
\hspace{-0.5cm}\underset{\mathbf{V},\mathbf{Q}^{\mathrm{TX}}, \mathbf{Q}^{\mathrm{RX}}}{\mathrm{max}}
&&\hspace{-0.5cm}\gamma_s(\mathbf{V},\hspace{-0.05cm} \mathbf{Q}^{\mathrm{TX}},\hspace{-0.05cm} \mathbf{Q}^{\mathrm{RX}}),\\
\text{s.t.}\qquad&&\hspace{-0.5cm}R_{\mathrm{Sec}}(\mathbf{V},\hspace{-0.05cm} \mathbf{Q}^{\mathrm{TX}})\hspace{-0.05cm}\geq\hspace{-0.05cm} \bar{R}_{\mathrm{Sec}},\\
&&\hspace{-0.5cm}R_{i}^{\mathrm{TX}}(\mathbf{V}_{i},\hspace{-0.05cm} \mathbf{Q}^{\mathrm{TX}}_{i})\hspace{-0.05cm}\leq\hspace{-0.05cm} \bar{C}_i^{\mathrm{TX}},\hspace{+0.05cm} \forall i \in \mathcal{N}_t,\\
&&\hspace{-0.5cm}R_{j}^{\mathrm{RX}}(\mathbf{V},\hspace{-0.05cm} \mathbf{Q}^{\mathrm{TX}},\hspace{-0.05cm} \mathbf{Q}_j^{\mathrm{RX}})\hspace{-0.05cm}\leq\hspace{-0.05cm} \bar{C}_j^{\mathrm{RX}},\hspace{+0.05cm} \forall j \in \mathcal{N}_r,\\
&&\hspace{-0.5cm}P_{i}(\mathbf{V}_{i},\hspace{-0.05cm} \mathbf{Q}^{\mathrm{TX}}_{i})\hspace{-0.05cm}\leq\hspace{-0.05cm} \bar{P}_i,\hspace{+0.05cm} \forall i \in \mathcal{N}_t,\\
&&\hspace{-0.5cm}\mathbf{V}_k \succeq 0, \mathrm{rank}(\mathbf{V}_k) = 1,
\end{eqnarray}
\end{subequations}
where $\mathbf{V}$$=$$\{\mathbf{V}_{k}\}_{k\in \mathcal{K}}$. The problem (\ref{p2}) is not convex due to the non-convextiy of (9a)-(9d) and (9f).

To address the non-convexity of (9), the objective function is a linear-fractional function of ${\mathbf{V}}$, ${\mathbf{Q}^{\mathrm{TX}}}$ and  ${\mathbf{Q}^{\mathrm{RX}}}$, which can be transformed into a linear function \cite{r11}. By introducing the substitute variables $\mathbf{\Gamma}_{k} \triangleq \mathbf{V}_{k}/\Xi(\mathbf{V},\mathbf{Q}^{\mathrm{TX}},\mathbf{Q}^{\mathrm{RX}})$,
$\mathbf{\Omega}_i^{\mathrm{TX}} \triangleq \mathbf{Q}_{i}^{\mathrm{TX}}/\Xi(\mathbf{V},\mathbf{Q}^{\mathrm{TX}},\mathbf{Q}^{\mathrm{RX}})$,
$\mathbf{\Omega}_j^{\mathrm{RX}} \triangleq \mathbf{Q}_{j}^{\mathrm{RX}}/\Xi(\mathbf{V},\mathbf{Q}^{\mathrm{TX}},\mathbf{Q}^{\mathrm{RX}})$
and $z \triangleq 1/\Xi(\mathbf{V},\mathbf{Q}^{\mathrm{TX}},\mathbf{Q}^{\mathrm{RX}})$,
with $\Xi(\mathbf{V},\mathbf{Q}^{\mathrm{TX}},\mathbf{Q}^{\mathrm{RX}}) = \mathrm{tr}(\mathbf{C}(\mathbf{V}^H+\mathbf{Q}^{\mathrm{TX}})\mathbf{C}^H) +\sum_{j=1}^{N_R}N_A \sigma_{y,j}^2+\sum_{j=1}^{N_R}\mathrm{tr}(\mathbf{Q}_j^{\mathrm{RX}})$, the problem (9) can be transformed into
\vspace{-0.2cm}
\begin{subequations}
\begin{eqnarray}
&&\hspace{-1.2cm}\underset{\mathbf{\Gamma},\mathbf{\Omega}^{\mathrm{TX}},\mathbf{\Omega}^{\mathrm{RX}}, z}{\mathrm{max}} \mathrm{tr}(\hspace{-0.05cm}\mathbf{G}(\mathbf{\Gamma}\hspace{-0.05cm}+\hspace{-0.05cm}\mathbf{\Omega}^{\mathrm{TX}})\mathbf{G}^H\hspace{-0.05cm})\label{P2,a}\\
\hspace{-1cm}\text{s.t.}&&\hspace{-0.6cm}\tilde{R}_{\mathrm{Sec,k}}(\mathbf{\Gamma},\hspace{-0.05cm} \mathbf{\Omega}^{\mathrm{TX}},\hspace{-0.05cm} z)\hspace{-0.05cm}\geq\hspace{-0.05cm} \bar{R}_{\mathrm{Sec}},\hspace{+0.05cm} \forall k \in \mathcal{K}\label{P2,b}\\
&&\hspace{-0.7cm}\log\hspace{-0.05cm}\det(\mathbf{\Gamma}_i\hspace{-0.05cm}+\hspace{-0.05cm}\mathbf{\Omega}^{\mathrm{TX}}_{i})\hspace{-0.05cm}-\hspace{-0.05cm}\log\hspace{-0.05cm}\det(\mathbf{\Omega}^{\mathrm{TX}}_{i})\hspace{-0.05cm}\leq\hspace{-0.05cm} \bar{C}_i^{\mathrm{TX}}\hspace{-0.05cm}, \forall i \in \mathcal{N}_t, \label{P2,c}\\
&&\hspace{-0.7cm}\log\hspace{-0.05cm}\det(\hspace{-0.05cm}(\hspace{-0.05cm}\mathbf{G}_j\hspace{-0.05cm}+\hspace{-0.05cm}\mathbf{C}_j\hspace{-0.05cm})(\hspace{-0.05cm}\mathbf{\Gamma}\hspace{-0.05cm}+\hspace{-0.05cm}\mathbf{\Omega}^{\mathrm{TX}}\hspace{-0.05cm})(\hspace{-0.05cm}\mathbf{G}_j+\mathbf{C}_j\hspace{-0.05cm})^H \nonumber\\
&&\hspace{-0.7cm}+ \hspace{-0.05cm}z\sigma_{y,j}^2\mathbf{I}_{N_A}\hspace{-0.05cm}+\hspace{-0.05cm}\mathbf{\Omega}_j^{\mathrm{RX}})\hspace{-0.05cm}-\hspace{-0.05cm}\log\hspace{-0.05cm}\det(\hspace{-0.05cm}\mathbf{\Omega}_j^{\mathrm{RX}}\hspace{-0.05cm})\hspace{-0.05cm}\leq\hspace{-0.05cm} \bar{C}_j,\hspace{+0.05cm} \forall j \in \mathcal{N}_r,\label{P2,d}\\
&&\hspace{-0.7cm}\mathrm{tr}(\mathbf{\Gamma}_i\hspace{-0.05cm}+\hspace{-0.05cm}\mathbf{\Omega}^{\mathrm{TX}}_{i})\hspace{-0.05cm}-\hspace{-0.05cm}z\bar{P}_i\hspace{-0.05cm}\leq\hspace{-0.05cm} 0,\hspace{+0.05cm} \forall i \in \mathcal{N}_t,\label{P2,e}\\
&&\hspace{-0.7cm}\mathbf{\Gamma}_k \succeq 0, \mathrm{rank}(\mathbf{\Gamma}_k)\hspace{-0.05cm} =\hspace{-0.05cm} 1,\hspace{+0.05cm} \forall k \in \mathcal{K}\label{P2,f}\\
&&\hspace{-0.9cm}\mathrm{tr}(\hspace{-0.05cm}\mathbf{C}(\mathbf{\Gamma}\hspace{-0.05cm}+\hspace{-0.05cm}\mathbf{\Omega}^{\mathrm{TX}}\hspace{-0.05cm})\mathbf{C}^H\hspace{-0.05cm})\hspace{-0.02cm}\hspace{-0.05cm}+\hspace{-0.05cm}z\hspace{-0.05cm}\sum_{j=1}^{N_R}\hspace{-0.05cm}N_A \hspace{-0.01cm}\sigma_{y,j}^2\hspace{-0.05cm}+\hspace{-0.1cm}\sum_{j=1}^{N_R}\hspace{-0.05cm}\mathrm{tr}(\hspace{-0.05cm}\mathbf{\Omega}_j^{\mathrm{RX}}\hspace{-0.05cm}) \hspace{-0.05cm}=\hspace{-0.05cm} 1, \label{P2,g}
\end{eqnarray}
\end{subequations}
where we have defined 
\vspace{-0.2cm}
\begin{eqnarray}
&&\hspace{-1cm}\tilde{R}_{\mathrm{Sec,k}}(\mathbf{\Gamma}, \mathbf{\Omega}^{\mathrm{TX}}, z) \nonumber\\
&&\hspace{-1cm}=\log\hspace{-0.05cm}\left(\hspace{-0.05cm}1\hspace{-0.05cm}+\hspace{-0.05cm}\frac{\mathbf{h}_k^H\mathbf{\Gamma}_k\mathbf{h}_k}{\mathbf{h}_k^H\hspace{-0.1cm}\left(\hspace{-0.05cm}\sum_{k'\in \mathcal{K} \backslash k}\mathbf{\Gamma}_{k'}\hspace{-0.05cm}+\hspace{-0.05cm}\mathbf{\Omega}^{\mathrm{TX}}\hspace{-0.05cm}\right)\hspace{-0.05cm}\mathbf{h}_k\hspace{-0.05cm}+\hspace{-0.05cm}z\sigma_k^2}\hspace{-0.05cm}\right) \nonumber\\
&&\hspace{-1cm} \quad -\log\hspace{-0.05cm}\left(\hspace{-0.05cm}1\hspace{-0.05cm}+\hspace{-0.05cm}\frac{\mathbf{h}_E^H\mathbf{\Gamma}_k\mathbf{h}_E}{\mathbf{h}_E^H\hspace{-0.1cm}\left(\hspace{-0.05cm}\sum_{k'\in \mathcal{K} \backslash k}\mathbf{\Gamma}_{k'}\hspace{-0.05cm}+\hspace{-0.05cm}\mathbf{\Omega}^{\mathrm{TX}}\hspace{-0.05cm}\right)\mathbf{h}_E\hspace{-0.05cm}+\hspace{-0.05cm}z\sigma_E^2}\hspace{-0.05cm}\right). \nonumber
\end{eqnarray}
Furthermore, since (\ref{P2,b}), (\ref{P2,c}) and (\ref{P2,d}) are differences of convex (DC) functions, and (\ref{P2,f}) represents the rank-1 constraint, we adopt the MM method \cite{r12} and rank relaxation \cite{r13}. The overall algorithm to solve (10) is summarized in Algorithm 1. At the $n$th iteration of the loop in Algorithm 1, the variables $\mathbf{\Gamma}^{(n)}$, $\mathbf{\Omega}^{\mathrm{TX}, (n)}$ $\mathbf{\Omega}^{\mathrm{RX}, (n)}$ and $z^{(n)}$ can be obtained by solving an approximation of problem (10) with respect to $\mathbf{\Gamma}^{(n-1)}$, $\mathbf{\Omega}^{\mathrm{TX}, (n-1)}$ $\mathbf{\Omega}^{\mathrm{RX}, (n-1)}$ and $z^{(n-1)}$.

\subsection{Proposed Algorithm}
In the following, we provide the details of Algorithm 1. To simplify the notation, we define the optimization variable set as $\mathbf{\Theta}$$=$$\{\mathbf{\Gamma}, \mathbf{\Omega}^{\mathrm{TX}}, \mathbf{\Omega}^{\mathrm{RX}}, z\}$. Firstly, for the non-convex constraint (\ref{P2,b}), we derive the tight lower bound, $\bar{R}_{\mathrm{Sec}, k}(\mathbf{\Theta}\,|\,\mathbf{\Theta}^{(n-1)})$, by using the first-order Taylor approximation \cite{r7}, which is given as
\vspace{-0.1cm}
\begin{eqnarray}
&&\hspace{-1cm}\check{R}_{\mathrm{Sec}, k}\hspace{-0.05cm}(\hspace{-0.05cm}\mathbf{\Theta}\hspace{-0.05cm}\,|\,\hspace{-0.05cm}\mathbf{\Theta}^{(n-1)}\hspace{-0.05cm}) \hspace{-0.05cm}= \hspace{-0.05cm}\log\hspace{-0.08cm}\left(\hspace{-0.1cm}\mathbf{h}_k^H\hspace{-0.1cm}\left(\hspace{-0.05cm}\sum_{k'\in \mathcal{K}}\hspace{-0.05cm}\mathbf{\Gamma}_{k'}\hspace{-0.05cm}+\hspace{-0.05cm}\mathbf{\Omega}^{\mathrm{TX}}\hspace{-0.1cm}\right)\hspace{-0.05cm}\mathbf{h}_k \hspace{-0.05cm}+\hspace{-0.05cm} z\sigma_k^2\hspace{-0.05cm}\right) \nonumber \\ 
&&\hspace{-1cm}+\log\hspace{-0.08cm}\left(\hspace{-0.1cm}\mathbf{h}_E^H\hspace{-0.1cm}\left(\hspace{-0.05cm}\sum_{k'\in \mathcal{K} \backslash k}\hspace{-0.05cm}\hspace{-0.1cm}\mathbf{\Gamma}_{k'}\hspace{-0.05cm} +\hspace{-0.05cm}\mathbf{\Omega}^{\mathrm{TX}}\hspace{-0.1cm}\right)\hspace{-0.1cm}\mathbf{h}_E\hspace{-0.05cm}+\hspace{-0.05cm}z\sigma_E^2\hspace{-0.1cm}\right) \hspace{-0.1cm}-\hspace{-0.1cm}f(\hspace{-0.05cm}\mathbf{\Theta}\hspace{-0.05cm}\,|\,\hspace{-0.05cm}\mathbf{\Theta}^{(n-1)}\hspace{-0.05cm}),
\end{eqnarray}
with 
\vspace{-0.2cm}
\begin{eqnarray}
&&\hspace{-0.7cm} f(\mathbf{\Theta}\,|\,\mathbf{\Theta}^{(n-1)}) \nonumber\\
&&\hspace{-0.7cm}\triangleq \log\hspace{-0.05cm}\left(\hspace{-0.05cm}\mathbf{h}_k^H\hspace{-0.05cm}\left(\hspace{-0.05cm}\sum_{k'\in \mathcal{K} \backslash k}\hspace{-0.05cm}\mathbf{\Gamma}_{k'}^{(n-1)}\hspace{-0.05cm}+\hspace{-0.05cm}\mathbf{\Omega}^{\mathrm{TX}, (n-1)}\hspace{-0.05cm}\right)\hspace{-0.05cm}\mathbf{h}_k\hspace{-0.05cm}+\hspace{-0.05cm}z^{(n-1)}\sigma_k^2\hspace{-0.05cm}\right) \nonumber\\
&&\hspace{-0.5cm}+\log\hspace{-0.05cm}\left(\hspace{-0.05cm}\mathbf{h}_E^H\hspace{-0.05cm}\left(\hspace{-0.05cm}\sum_{k'\in \mathcal{K}}\hspace{-0.05cm}\mathbf{\Gamma}_{k'}^{(n-1)}\hspace{-0.05cm}+\hspace{-0.05cm}\mathbf{\Omega}^{\mathrm{TX}, (n-1)}\hspace{-0.05cm}\right)\hspace{-0.05cm}\mathbf{h}_E\hspace{-0.05cm}+\hspace{-0.05cm}z^{(n-1)}\sigma_E^2\hspace{-0.05cm}\right)\nonumber\\
&&\hspace{-0.5cm}+\frac{1}{\ln 2}\hspace{-0.1cm}\left[\hspace{-0.05cm}\frac{\mathbf{h}_k^H\hspace{-0.05cm}\left(\hspace{-0.05cm}\sum_{k'\in \mathcal{K}\backslash k}\hspace{-0.05cm}\mathbf{\Gamma}_{k'}\hspace{-0.05cm}+\hspace{-0.05cm}\mathbf{\Omega}^{\mathrm{TX}}\hspace{-0.05cm}\right)\hspace{-0.05cm}\mathbf{h}_k\hspace{-0.05cm} +\hspace{-0.05cm} z}{\mathbf{h}_k^H\hspace{-0.05cm}\left(\hspace{-0.05cm}\sum_{k'\in \mathcal{K}\backslash k}\hspace{-0.05cm}\mathbf{\Gamma}_{k'}^{(n-1)}\hspace{-0.05cm}+\hspace{-0.05cm} \mathbf{\Omega}^{\mathrm{TX}, (n-1)}\hspace{-0.05cm}\right)\hspace{-0.05cm}\mathbf{h}_k\hspace{-0.05cm} + z^{(n-1)}}\hspace{-0.05cm} -\hspace{-0.05cm} 1\hspace{-0.05cm}\right] \nonumber\\
&&\hspace{-0.5cm}+\frac{1}{\ln 2}\hspace{-0.1cm}\left[\hspace{-0.05cm}\frac{\mathbf{h}_E^H\hspace{-0.05cm}\left(\hspace{-0.05cm}\sum_{k'\in \mathcal{K}}\hspace{-0.05cm}\mathbf{\Gamma}_{k'}\hspace{-0.05cm}+\hspace{-0.05cm}\mathbf{\Omega}^{\mathrm{TX}}\hspace{-0.05cm}\right)\hspace{-0.05cm}\mathbf{h}_E\hspace{-0.05cm} + z}{\mathbf{h}_E^H\hspace{-0.05cm}\left(\hspace{-0.05cm}\sum_{k'\in \mathcal{K}}\hspace{-0.05cm}\mathbf{\Gamma}_{k'}^{(n-1)}\hspace{-0.05cm}+\hspace{-0.05cm} \mathbf{\Omega}^{\mathrm{TX}, (n-1)}\hspace{-0.05cm}\right)\hspace{-0.05cm}\mathbf{h}_E\hspace{-0.05cm} + z^{(n-1)}}\hspace{-0.05cm} -\hspace{-0.05cm} 1\hspace{-0.05cm}\right]. \nonumber
\end{eqnarray}

Similarly, the tight upper bounds for (\ref{P2,c}) and (\ref{P2,d}) can be obtained by the first-order Taylor approximation as
\begin{eqnarray}
&&\hspace{-0.9cm}\tilde{R}_{i}^{\mathrm{TX}}\hspace{-0.05cm}(\hspace{-0.05cm}\mathbf{\Theta}\hspace{-0.05cm}\,|\,\hspace{-0.05cm}\mathbf{\Theta}^{(n-1)}\hspace{-0.05cm})\nonumber\\
&&\hspace{-0.9cm}=\hspace{-0.05cm}g\hspace{-0.05cm}\left(\hspace{-0.05cm}\mathbf{\Gamma}_{i}^{(n-1)}\hspace{-0.05cm}+\hspace{-0.05cm}\mathbf{\Omega}^{\mathrm{TX}, (n-1)}_{i},\mathbf{\Gamma}_{i}\hspace{-0.05cm}+\hspace{-0.05cm}\mathbf{\Omega}^{\mathrm{TX}}_{i}\hspace{-0.05cm}\right)\hspace{-0.05cm}-\hspace{-0.05cm}\log\hspace{-0.05cm}\det(\mathbf{\Omega}^{\mathrm{TX}}_{i}),\\
&&\hspace{-0.9cm}\tilde{R}_j^{\mathrm{RX}}\hspace{-0.05cm}(\hspace{-0.05cm}\mathbf{\Theta}\hspace{-0.05cm}\,|\,\hspace{-0.05cm}\mathbf{\Theta}^{(n-1)}\hspace{-0.05cm})\hspace{-0.05cm}=\hspace{-0.05cm} g\hspace{-0.05cm}\left(\hspace{-0.05cm}a(\hspace{-0.05cm}\mathbf{\Theta}^{(n-1)}\hspace{-0.05cm}), a(\hspace{-0.05cm}\mathbf{\Theta}\hspace{-0.05cm})\hspace{-0.1cm}\right)\hspace{-0.05cm}-\hspace{-0.05cm}\log\hspace{-0.05cm}\det(\mathbf{\Omega}_j^{\mathrm{RX}}),
\end{eqnarray}
where $a(\hspace{-0.05cm}\mathbf{\Theta}\hspace{-0.05cm}) \triangleq  (\mathbf{G}_j+\mathbf{C}_j)(\mathbf{W}\mathbf{W}^H + \mathbf{Q}^{\mathrm{TX}})(\mathbf{G}_j+\mathbf{C}_j)^H+\sigma_{y,j}^2\mathbf{I}_{N_A}+\mathbf{Q}_j^{\mathrm{RX}}$
and $g(\mathbf{X}_1,\mathbf{X}_2)\triangleq \log\hspace{-0.05cm}\det(\mathbf{X}_1)+\mathrm{tr}(\mathbf{X}_1^{-1}(\mathbf{X}_2-\mathbf{X}_1))/\ln 2$ \cite{r7}. 
The problem (14) in Algorithm 1 is convex and can be solved by interior point method or standard tool, e.g., CVX \cite{r11}.
\begin{algorithm}
\small
\caption{Algorithm for the maximization of the sensing SINR in Secure Cell-free ISAC Systems}
\begin{algorithmic}
\State \textbf{1. Initialize:} $\mathbf{\Theta}^{(0)}$ 
\State \textbf{2. Repeat} (Until $\|\mathbf{\Theta}^{(n)}-\mathbf{\Theta}^{(n-1)}\|\leq \epsilon$ with a threshold $\epsilon$)\\
\hspace{+0.2cm}$n \gets n+1$\\
\hspace{+0.2cm}Update $\mathbf{\Theta}^{(n)} \gets \mathbf{\Theta}$ obtained by solving the following problem
\begin{subequations}
\begin{eqnarray}
&&\hspace{-0.8cm}\underset{\mathbf{\Theta}}{\mathrm{max}} \hspace{0.1cm}\mathrm{tr}(\hspace{-0.05cm}\mathbf{G}(\mathbf{\Gamma}\hspace{-0.05cm}+\hspace{-0.05cm}\mathbf{\Omega}^{\mathrm{TX}})\mathbf{G}^H\hspace{-0.05cm}), \\
&&\hspace{-0.6cm}\text{s.t.} \hspace{+0.1cm}\check{R}_{\mathrm{Sec}, k}\hspace{-0.05cm}(\hspace{-0.05cm}\mathbf{\Theta}\hspace{-0.05cm}\,|\,\hspace{-0.05cm}\mathbf{\Theta}^{(n-1)}\hspace{-0.05cm}),\\
&&\hspace{-0.1cm} \tilde{R}_{i}^{\mathrm{TX}}\hspace{-0.05cm}(\hspace{-0.05cm}\mathbf{\Theta}\hspace{-0.05cm}\,|\,\hspace{-0.05cm}\mathbf{\Theta}^{(n-1)}\hspace{-0.05cm})\leq\bar{C}_i^{\mathrm{TX}}, \forall i \in \mathcal{N}_t,\\
&&\hspace{-0.1cm} \tilde{R}_j^{\mathrm{RX}}\hspace{-0.05cm}(\hspace{-0.05cm}\mathbf{\Theta}\hspace{-0.05cm}\,|\,\hspace{-0.05cm}\mathbf{\Theta}^{(n-1)}\hspace{-0.05cm}) \leq \bar{C}_j^{\mathrm{RX}},\hspace{+0.05cm}\forall j \in \mathcal{N}_r,\\
&&\hspace{-0.1cm}\mathrm{tr}(\mathbf{\Gamma}_i\hspace{-0.05cm}+\hspace{-0.05cm}\mathbf{\Omega}^{\mathrm{TX}}_{i})\hspace{-0.05cm}-\hspace{-0.05cm}z\bar{P}_i\hspace{-0.05cm}\leq\hspace{-0.05cm} 0,\hspace{+0.05cm} \forall i \in \mathcal{N}_t,\\
&&\hspace{-0.1cm}\mathbf{\Gamma}_k \succeq 0, \hspace{+0.05cm}  \forall k \in \mathcal{K},\\
&&\hspace{-1.5cm}\mathrm{tr}\hspace{-0.02cm}(\hspace{-0.02cm}\mathbf{C}\hspace{-0.02cm}(\hspace{-0.02cm}\mathbf{\Gamma}\hspace{-0.05cm}+\hspace{-0.05cm}\mathbf{\Omega}^{\mathrm{TX}})\hspace{-0.02cm}\mathbf{C}^H\hspace{-0.02cm})\hspace{-0.05cm}+\hspace{-0.05cm}z\hspace{-0.05cm}\sum_{j=1}^{N_R}\hspace{-0.02cm}N_A\hspace{-0.02cm} \sigma_{y,j}^2\hspace{-0.05cm}+\hspace{-0.05cm}\sum_{j=1}^{N_R}\hspace{-0.02cm}\mathrm{tr}\hspace{-0.02cm}(\hspace{-0.02cm}\mathbf{\Omega}_j^{\mathrm{RX}}\hspace{-0.02cm})\hspace{-0.05cm}=\hspace{-0.05cm}1,
\end{eqnarray}
\end{subequations} 
\State \textbf{3. Calculate:} $\mathbf{Q}^{\mathrm{TX}} = \mathbf{\Omega}^{\mathrm{TX}}/z$, $\mathbf{Q}^{\mathrm{RX}} = \mathbf{\Omega}^{\mathrm{RX}}/z$ and $\mathbf{w}_k = \zeta_k\sqrt{\lambda_k}\mathbf{v}_k$, where $\mathbf{v}_k$ and $\lambda_k$ are eigenvector and eigenvalue of $\mathbf{V}_k = \mathbf{\Gamma}_k/z$, and $\zeta_k$ is the scaling factor.
\State \textbf{4. Output:} $\{\mathbf{w}_k\}_{\forall k \in \mathcal{K}}$, $\mathbf{\Omega}^{\mathrm{TX}}$, $\mathbf{\Omega}^{\mathrm{RX}}$
\end{algorithmic}
\end{algorithm}

\subsection{Convergence and Complexity Analysis}
In this section, we analyze the convergence and complexity for Algorithm 1. Let $\chi$ denote the feasible set of problem (10), and let $\chi(\nu^{(n-1)})$ and $\nu^{(n)}$ denote the feasible set and point of problem (14) for the $n$th iteration of Algorithm 1. By the tight bound of the DC structure for (14b), (14c) and (14d), it is ensured that $\nu^{(n)}$$\in$$\chi(\nu^{(n)}) $$\subseteq$$\chi$. Also, the problem (14) converges when the step size is defined as $\phi^{(\beta)}$, where $\beta$ is the number of iterations of (14), and the sequence of $\phi^{(\beta)}$ is chosen as $\phi^{(\beta)}$$\in$$(0,1]$, $\phi^{(\beta)}$$\rightarrow$$0$, and $\sum_{\beta} \phi^{(\beta)}$$\rightarrow$$\infty$. Accordingly, for every $n$, the inequality $\tilde{\gamma}_s(\mathbf{\Theta}^{(0)})$$\leq$$ \tilde{\gamma}_s(\mathbf{\Theta}^{(n)})$$\leq$$ \tilde{\gamma}_s(\mathbf{\Theta}^{(n+1)})$$\leq$$ \tilde{\gamma}_s(\mathbf{\Theta}^{(\infty)})$ holds,  which guarantees the convergence of Algorithm 1 \cite{r14}, with $\tilde{\gamma}_s(\mathbf{\Theta}) \triangleq \mathrm{tr}(\hspace{-0.05cm}\mathbf{G}(\mathbf{\Gamma}\hspace{-0.05cm}+\hspace{-0.05cm}\mathbf{\Omega}^{\mathrm{TX}})\mathbf{G}^H\hspace{-0.05cm})$.

For the computational complexity analysis, we denote the number of iterations of Algorithm 1 as $\mathcal{I}$. In each iteration, problem (14) is solved using the well-known Primal-Dual Interior Point Method \cite{r15} to perform a Newton step update in each loop, whose the number of loops is approximately bounded by $\mathcal{O}(\sqrt{K+N_T+N_R})$. Here, since the computational complexity of each loop depends on the dominant size of $\mathbf{\Gamma}$, the computational complexity of a single Newton step update can be evaluated as $\mathcal{O}(K^{3}N_T^{6}N_A^{6})$, resulting from the matrix inversion and multiplication. Therefore, the overall computational complexity of Algorithm 1 can be calculated as $\mathcal{O}(\mathcal{I}K^{3}N_T^{6}N_A^{6}\sqrt{K+N_T+N_R})$.

\section{Numerical Results}
In this section, we evaluate the performance of the proposed algorithm via numerical results. For simplicity, Tx-RRH is denoted as TX and Rx-RRH as RX. By referring to \cite{r7}, we adopt equal power constraints for all TXs of $P$$=$$\{\bar{P}_i\}_{i \in \mathcal{N}_t}$ and the equal fronthaul link constraints of $C^{\text{TX}}$$=$$\{\bar{C}^{\text{TX}}_i\}_{i \in \mathcal{N}_t}$ and $C^{\text{RX}}$$=$$\{\bar{C}^{\text{RX}}_j\}_{j \in \mathcal{N}_r}$. The variance of the sensing channel and the clutter channel are set to $\sigma_{G, j,i}^2$$=$$0.001$, and $\sigma_{C, j, i}^2$$=$$0.001$, for all $\{i, j\}$$\in$$\mathcal{N}_t \times \mathcal{N}_r$, and the Rician factor is set as $R$$=$$5$. The AWGN variances is set to 0.1, and the total sensing duration is set to 30 symbol times \cite{r4, r5, r10}.

First, we analyze the receiver operating characteristic (ROC) curves shown in Fig 2. The ROC curves are generated by applying the optimum test detectors \cite{r10}, while varying the decision threshold. In Fig. 2, two TXs are positioned at (0,500) and (500,500), while two RXs are located at (0,250) and (500,250). The Eve's location is uniformly generated within the square region centered at (250,250) with a side length of 30 meters, while the users are uniformly placed within same-size square region centered at (200,200) and (300,300) respectively. It is observed that as the secrecy rate constraint increases, the sensing performance decreases, which represents the trade-off between communications and sensing. This is because the stronger secrecy rate constraint makes it difficult to steer the beamforming vector towards Eve, which in turn worsens the sensing performance.

In Fig 3, we compare the performance with reference schemes based on the fronthaul capacity $C^{\text{RX}}$. We consider two reference schemes; distributed sensing, where RXs detect the target and the CU decides via majority rule, and random beamforming, which performs randomly beamforming ignoring channel information. The performance metric of the sensing accuracy is defined as $P_{sa}$$=$$(P_{de} + P_{fa})/2$, where $P_{fa}= 0.1$. The proposed algorithm exhibits improved sensing capability as receiver and transmitter fronthaul capacity increases due to reduced compression losses. Moreover, when comparing the proposed algorithm to the distributed sensing scheme, the proposed algorithm outperforms when the fronthaul capacity is sufficiently large ($C^{\text{RX}} \geq 3$). This is due to the fact that with sufficient fronthaul capacity, the benefits of joint processing outweigh the degradation caused by compression.

\begin{figure}[t]
    \centering
    \includegraphics[width= 7.5cm]{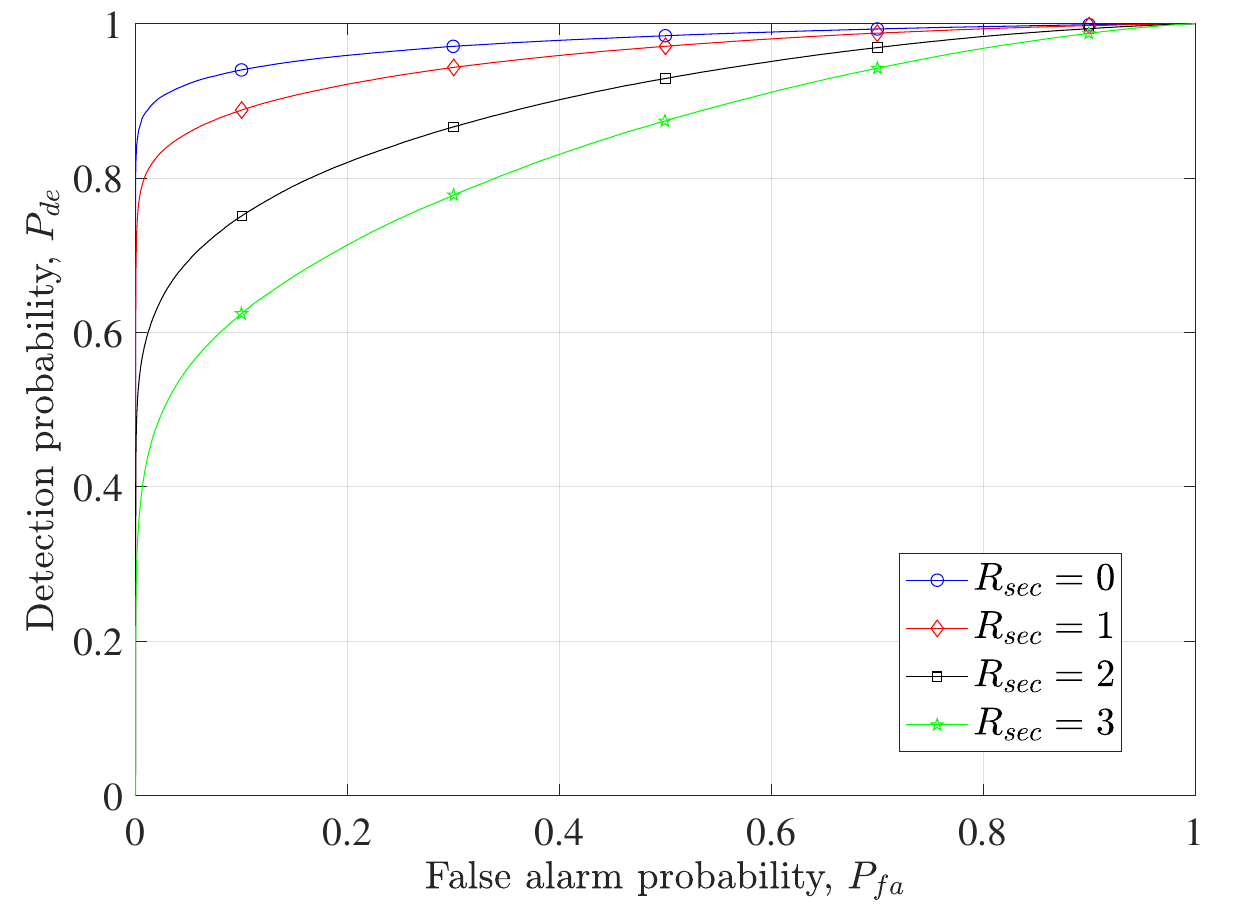}
    \vspace{-0.1cm}
    \caption{ROC curves with different secrecy rate constraints ($N_{A} = 3$, $P = 6$, $N_{T}=2$, $N_{R}=2$, $K = 2$)}
    \label{fig:sim1}
    \vspace{-0.35cm}
\end{figure}

\begin{figure}[t]
    \centering
    \includegraphics[width= 7.5cm]{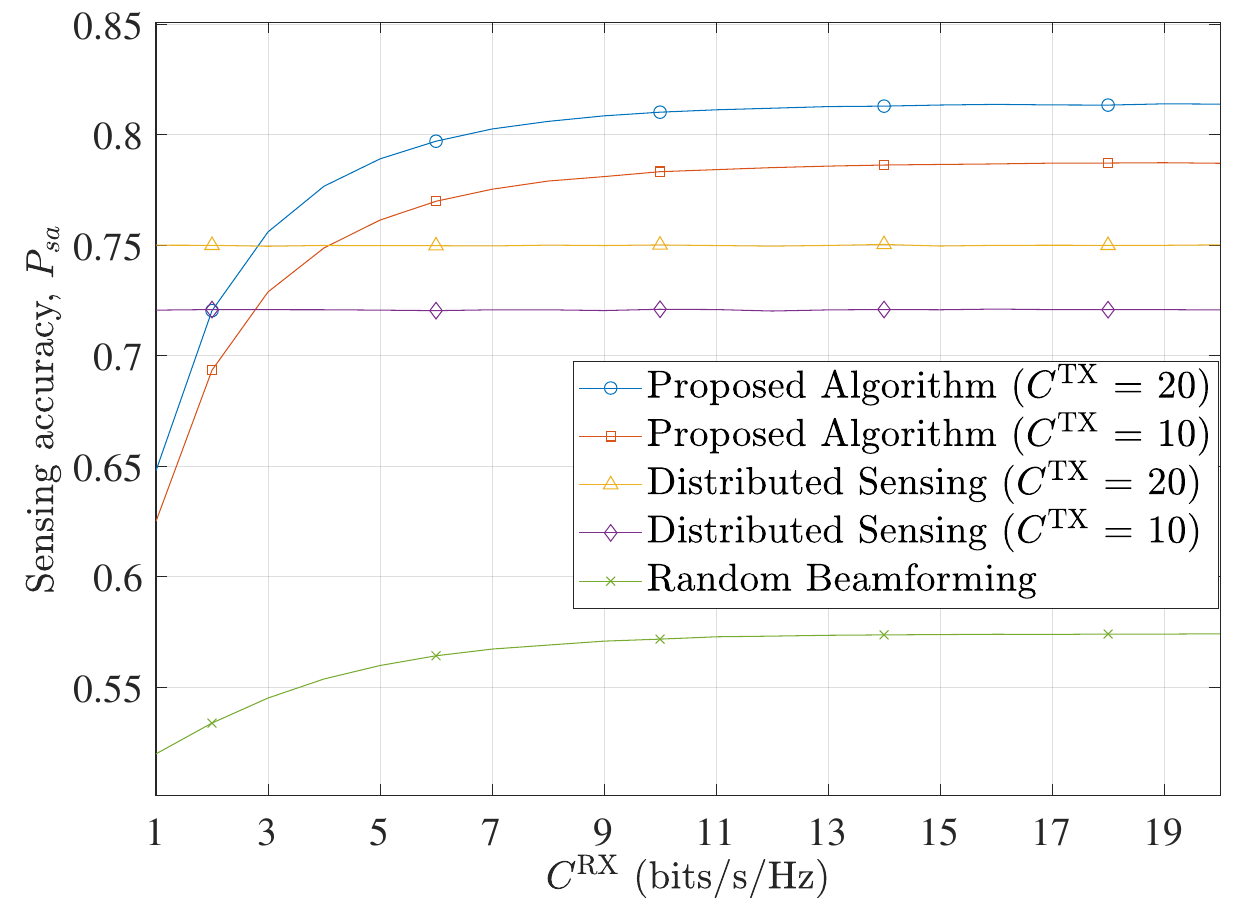}
    \vspace{-0.1cm}
    \caption{Sensing accuracy versus fronthaul rate constraint at the RX side ($N_{A} = 2$, $P = 5$, $N_{T}=2$, $N_{R}=2$, $K = 2$)}
    \label{fig:sim2}
    \vspace{-0.35cm}
\end{figure}

Fig. 4 illustrates the sensing performance based on the user's position and the number of TXs. For the better understanding, we consider only the LOS component in communication channel. The positions of the TXs, the user, and Eve are shown in Fig. 4(b), where the user is placed on a circle with a radius of 400 meters from the center. The sensing SINR, calculated based on the user’s position in Fig. 4(b), is shown in Fig. 4(a), where the discontinuous points are the locations not to satisfy the given constraints. It is noticed that for the smaller number of the TXs, the discontinuities occur over the wider range. This is because, the beam resolution to distinguish between angles becomes difficult as the user is located farther away. In other words, adding the TXs facilitates finding an appropriate transmit beamforming vector that satisfies the secrecy rate constraint due to the spatial diversity. Additionally, in Case 4, the sensing SINR becomes uniform with a consistent level at all the user's location, which is the benefit of the cell-free architecture for the secure ISAC systems.

\begin{figure}[t]
    \centering
    \includegraphics[width=8.5cm]{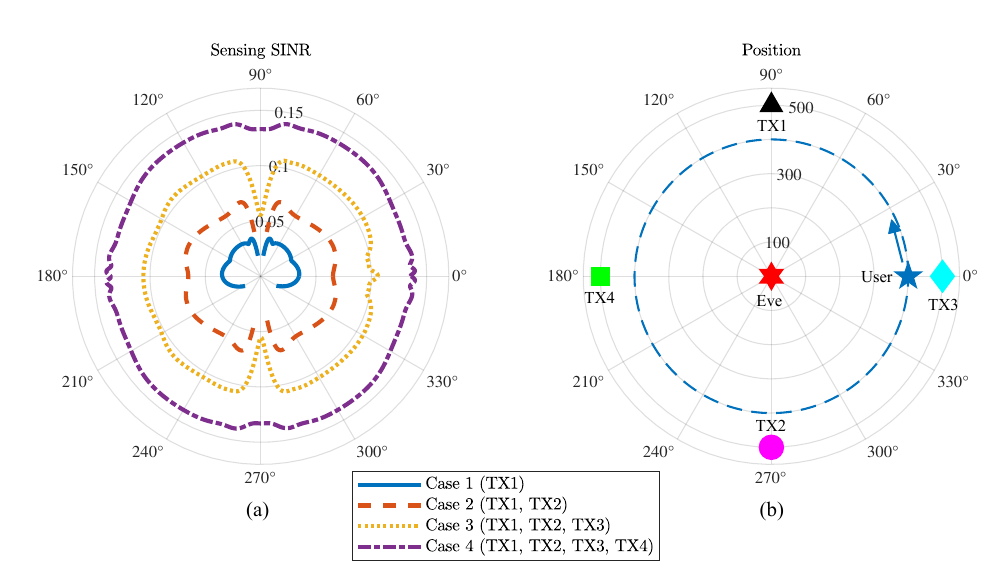}
    \caption{Sensing SINR versus user position across various cases ($N_{A} = 6$, $P = 5$, $N_{R}=1$)}
    \label{fig:sim3}
    \vspace{-0.5cm}
\end{figure}
\section{Conclusions}
This correspondence proposes the joint design of the transmit beamforming and transmit/receive quantizers to maximize the sensing performance under the constraints on secrecy rate, fronthaul rate and maximum power in a secure cell-free ISAC system. Numerical results verify that the proposed algorithm provides superior sensing performance compared to the reference schemes that becomes more emphasized with the more fronthaul capacity and the more RRHs. As future works, we can extend to explore the robust design against the channel uncertainty and near-field scenario.

\bibliographystyle{IEEEtran}
\bibliography{ref}


%



\ifCLASSOPTIONcaptionsoff
  \newpage
\fi

\end{document}